\documentclass[sigplan, authorversion]{acmart}

\pdfoutput=1

\copyrightyear{2019}
\acmYear{2019}
\setcopyright{acmlicensed}
\acmConference[ASPLOS '19]{2019 Architectural Support for Programming Languages and Operating Systems}{April 13--17, 2019}{Providence, RI, USA}
\acmBooktitle{2019 Architectural Support for Programming Languages and Operating Systems (ASPLOS '19), April 13--17, 2019, Providence, RI, USA}
\acmPrice{15.00}
\acmDOI{10.1145/3297858.3304062}
\acmISBN{978-1-4503-6240-5/19/04}

\usepackage[normalem]{ulem}

\usepackage{fancyhdr}
\usepackage{microtype}
\usepackage{flushend}
\usepackage{ifthen}
\usepackage{textcomp}
\usepackage[vlined,ruled]{algorithm2e}
\usepackage{verbatim}
\usepackage{textgreek}
\usepackage{tikz}
\usetikzlibrary{shadows,arrows}
\usetikzlibrary{calc}
\usepackage{commath}

\usepackage{booktabs}

\usepackage{paralist}
\usepackage{hyperref}

\usepackage[shortcuts]{extdash}

\providecommand{\SetAlgoVlined}{\SetVline}
\providecommand{\DontPrintSemicolon}{\dontprintsemicolon}

\newboolean{blindreview}
\setboolean{blindreview}{false}

\newcommand{\void}[1]{}

\newcommand{\todo}[1]{\textbf{\color{red}[Todo: {\small #1}]}}
\renewcommand{\todo}[1]{}

\newcommand{\reftab}[1]{Table~\ref{tab:#1}}
\newcommand{\reffig}[1]{Figure~\ref{fig:#1}}
\newcommand{\refdef}[1]{Definition~\ref{def:#1}}

\newcommand{\refsec}[1]{Section~\ref{sec:#1}}
\newcommand{\refalg}[1]{Algorithm~\ref{alg:#1}}

\newcommand{\labeldef}[1]{\label{def:#1}}

\newcommand{\labelsec}[1]{\label{sec:#1}}

\newcommand{\labtab}[1]{\label{tab:#1}}

\newcommand{\labfig}[1]{\label{fig:#1}}

\newcommand{\labsec}[1]{\label{sec:#1}}
\newcommand{\labalg}[1]{\label{alg:#1}}

\usepackage{xspace}

\newcommand{\negvspace}{\vspace{-2mm}}

\newlength{\tmpTFS}

\newtheorem{definition}{Definition}

\newcommand{\microops}{{\textmu}ops\xspace}
\newcommand{\microop}{{\textmu}op\xspace}

\begin{document}

\title[uops.info]{uops.info: Characterizing Latency, Throughput, and Port Usage of Instructions on Intel Microarchitectures}

\author{Andreas Abel and Jan Reineke}
\email{{abel, reineke}@cs.uni-saarland.de}
\affiliation{%
  \institution{Saarland University\\
  Saarland Informatics Campus\\
  Saarbr\"ucken, Germany}
}

\begin{abstract}

Modern microarchitectures are some of the world's most complex man-made systems. 
As a consequence, it is increasingly difficult to predict, explain, let alone optimize the performance of software running on such microarchitectures.
As a basis for performance predictions and optimizations, we would need faithful models of their behavior, which are, unfortunately, seldom available. 

In this paper, we present the design and implementation of a tool to construct faithful models of the latency, throughput, and port usage of x86 instructions.
To this end, we first discuss common notions of instruction throughput and port usage, and introduce a more precise definition of latency that, in contrast to previous definitions, considers dependencies between different pairs of input and output operands.
We then develop novel algorithms to infer the latency, throughput, and port usage based on automatically-generated microbenchmarks that are more accurate and precise than existing work.
 
To facilitate the rapid construction of optimizing compilers and tools for performance prediction, the output of our tool is provided in a machine-readable format.
We provide experimental results for processors of all generations of Intel's Core architecture, i.e., from Nehalem to Coffee Lake, and discuss various cases where the output of our tool differs considerably from prior work.
\end{abstract}

\maketitle

\section{Introduction}

\todo{have any PC members done relevant work we could/should cite? maybe briefly skim their websites to identify potential reviewers; there's got to be more simulators and performance-analysis tools than we cite}

Developing tools that predict, explain, or even optimize the performance of software is challenging due to the complexity of today's microarchitectures. 
Unfortunately, this challenge is exacerbated by the lack of a precise documentation of their behavior.
While the high-level structure of modern microarchitectures is well-known and stable across multiple generations, lower-level aspects may differ considerably between microarchitecture generations and are generally not as well documented. 
An important aspect with a relatively strong influence on performance is how ISA instructions decompose into micro-operations (\microops); which ports these \microops may be executed on; and what their latencies are.

Knowledge of this aspect is required, for instance, to build precise performance-analysis tools like CQA~\cite{cqa14}, Kerncraft \cite{hammer15}, or llvm-mca~\cite{llvmmca}.
It is also useful when configuring cycle-accurate simulators like Zesto~\cite{Loh09}, gem5~\cite{Binkert11}, McSim+~\cite{Ahn13} or ZSim~\cite{sanchez13}.
Optimizing compilers, such as LLVM~\cite{llvm04} and GCC~\cite{gcc}, can profit from detailed instruction characterizations to generate efficient code for a specific microarchitecture.
Similarly, such knowledge is helpful when manually fine-tuning a piece of code for a specific processor. 

Unfortunately, information about the port usage, latency, and throughput of individual instructions at the required level of detail is hard to come by.
Intel's processor manuals~\cite{intelOptManual17} only contain latency and throughput data for a number of ``commonly-used instructions.''
They do not contain information on the decomposition of individual instructions into \microops, nor do they state the execution ports that these \microops can use.

The only way to obtain accurate instruction characterizations for many recent microarchitectures is thus to perform measurements using microbenchmarks. 
Such measurements are aided by the availability of performance counters that provide precise information on the number of elapsed cycles and the cumulative port usage of instruction sequences.
A relatively large body of work~\cite{Saavedra95,Thomborson00, Coleman01,Dongarra04,Yotov05a,Yotov06,Molka09,Babka09,Wong10,Abel13,Abel14,Hassan15,Mei17} uses microbenchmarks to infer properties of the memory hierarchy. 
Another line of work~\cite{Joshi08,Ganesan10,Ganesan11,Bertran12} uses automatically generated microbenchmarks to characterize the energy consumption of microprocessors.
Comparably little work~\cite{cqa14,fog17,granlund17,instlatx64} is targeted at instruction characterizations.
Furthermore, existing approaches, such as~\cite{fog17}\todo{``such as" suggests that there are more such approaches, and we should cite them! maybe we should also cite neighboring fields, such as approaches to characterize GPUs and memory hierarchies}, require significant manual effort to create the microbenchmarks and to interpret the results of the experiments.
Furthermore, its results are not always accurate and precise, as we will show later.\looseness=-1

In this paper, we develop a new approach that can automatically generate microbenchmarks in order to characterize the latency, throughput, and port usage of instructions on Intel Core CPUs in an accurate and precise manner.

Before describing our algorithms and their implementation, we first discuss common notions of instruction latency, throughput, and port usage.
For latency, we propose a new definition that, in contrast to previous definitions, considers dependencies between different pairs of input and output operands, which enables more accurate performance predictions.

We then develop algorithms that generate assembler code for microbenchmarks to measure the properties of interest for most x86 instructions.
Our algorithms take into account explicit and implicit dependencies, such as, e.g., dependencies on status flags.
Therefore, they require detailed information on the x86 instruction set.
We create a machine-readable XML representation of the x86 instruction set that contains all the information needed for automatically generating assembler code for each instruction. 
The relevant information is automatically extracted from the configuration files of Intel's \emph{x86 Encoder Decoder (XED)}~\cite{XED} library.

We have implemented our algorithms in a tool that we have successfully applied to all microarchitecture generations of Intel's Core architecture, i.e., from Nehalem to Coffee Lake.
In addition to running the generated microbenchmarks on the actual hardware, we have also implemented a variant of our tool that runs them on top of Intel IACA~\cite{IACA}.
IACA is a closed-source tool published by Intel that can statically analyze the performance of loop kernels on different Intel microarchitectures.
It is, however, updated only infrequently, and its results are not always accurate, as we will show later.

The output of our tool is available in a machine-readable format, so that it can be easily used to implement, e.g., simulators, performance prediction tools, or optimizing compilers.

Finally, we discuss several interesting insights obtained by comparing the results from our measurements with previously published data.
Our precise latency data, for example, uncovered previously undocumented differences between different microarchitectures.
It also explains discrepancies between previously published information.
Apart from that, we uncovered various errors in IACA, and inaccuracies in the manuals.
\todo{conclude with some positive note on its use?}


\section{Related Work}
In this section, we will describe existing sources of detailed instruction data for Intel microarchitectures.
We will first consider information provided by Intel directly, and then look at measurement-based approaches.

\subsection{Information provided by Intel}
Intel's \emph{Optimization Reference Manual}~\cite{intelOptManual17} contains a set of tables with latency and throughput data for ``commonly-used instructions.''
The tables are not complete; for some instructions, only throughput information is provided.
The manual does not contain detailed\todo{does it contain any such information?} information about the port usage of individual instructions.

IACA~\cite{IACA} is a closed-source tool developed by Intel that can statically analyze the performance of loop kernels on several microarchitectures (which can be different from the system that the tool is run on).
The tool generates a report which includes throughput and port usage data of the analyzed loop kernel.
By considering loop kernels with only one instruction, it is possible to obtain the throughput of the corresponding instruction.
However, it is, in general, not possible to determine the port bindings of the individual \microops this way.
Early versions of IACA were also able to analyze the latency of loop kernels; however, support for this was dropped in version 2.2.
IACA is updated only infrequently. 
Support for the Broadwell microarchitecture (which was released in 2014), for example, was added only in 2017.
There is currently no support for the two most recent microarchitectures, Kaby Lake and Coffee Lake, which were released in 2016 and 2017, respectively.

The instruction scheduling components of LLVM~\cite{llvm04} for the Sandy Bridge, Haswell, Broadwell, and Skylake microarchitecture were recently updated and extended with latency and port usage information that was, according to the commit message (\url{https://reviews.llvm.org/rL307529}), provided by the architects of these microarchitectures.
The resulting models are available in the LLVM repository.

\subsection{Measurement-based Approaches}
Agner Fog~\cite{fog17} provides lists of instruction latency, throughput, and port usage data for different x86 microarchitectures.
The data in the lists is not complete; e.g., latency data for instructions with memory operands is often missing.
The port usage information is sometimes inaccurate or imprecise; reasons for this are discussed in~\refsec{portUsage}.
The data is obtained using a set of test scripts developed by the author.
The output from these scripts has to be interpreted manually to build the instruction tables.

CQA~\cite{cqa14} is a performance analysis tool for x86 code that requires latency, throughput, and port usage data to build a performance model of a microarchitecture.
It includes a microbenchmark module that supports measuring the latency and throughput of many x86 instructions.\todo{can we say more about it?}
For non-supported instructions, the authors use Agner Fog's instruction tables~\cite{fog17}.
The paper briefly mentions that the module can also measure the number of \microops that are dispatched to execution ports using performance counters, but no further details are provided.

EXEgesis~\cite{EXEgesis} is a project that can create a machine-read-able list of instructions by parsing the PDF representation of Intel's Software Developer's Manual~\cite{intelDevManual}.
One of the goals of the project is also to infer latencies and \microop scheduling information for different instruction/microarchitecture pairs.

Granlund~\cite{granlund17} presents measured latency and throughput data for different x86 processors.
He considers only a relatively small subset of the x86 instruction set.\todo{can we briefly characterize how Granlund's approach works and what its deficiencies are, if any?}

AIDA64~\cite{AIDA} is a commercial, closed-source system information tool that can perform throughput and latency measurements of x86 instructions.
Measurement results for many processors obtained using AIDA64 are available at~\cite{instlatx64}.
\todo{possibly discuss EXEgesis by google, but there is very little documentation...}

\section{Background}
\labsec{background}

\subsection{Pipeline of Intel Core CPUs}
\labsec{pipeline}
\pgfdeclarelayer{background}
\pgfdeclarelayer{background1}
\pgfdeclarelayer{foreground}
\pgfsetlayers{background,background1,main,foreground}
 
\tikzstyle{nodeStyle} = [draw, text width=6cm,  minimum height=1.75em, text centered]
\tikzstyle{port} = [draw, fill={rgb,255:red,135; green,220; blue,170}, text width=0.75cm, font=\fontsize{6}{7.2}\sffamily, text centered]
\tikzstyle{FU} = [draw, fill={rgb,255:red,212; green,170; blue,0}, text width=5.0em, font=\fontsize{6}{7.2}\sffamily, rotate=90, text centered]
\tikzstyle{arrow} = [draw, thick, color=black!80, font=\footnotesize\sffamily]

\newcommand{\background}[7]{%
  \begin{pgfonlayer}{background}
    \path (#1.west |- #2.north)+(-1,0.4) node (a1) {};
    \path (#3.east |- #4.south)+(+0.4,#5) node (a2) {};
    \path[fill=#6, draw=black!50]
      (a1) rectangle (a2);
    \path let \p{x}=(a1), \p{y}=($(a1)!0.5!(a2)$) in (\x{x}, \y{y})+(0.5,0) node (u1)[rotate=90]
      {#7};
  \end{pgfonlayer}}
  
\begin{figure}
\centering
\begin{tikzpicture}[scale=.8,transform shape,font=\fontsize{11}{13.2}\sffamily] 
  \path node (nIC) [nodeStyle, fill={rgb,255:red,249; green,177; blue,166}] {Instruction Cache};
  \path (nIC.south)+(0.0,-1.0) node (nFD) [nodeStyle, fill={rgb,255:red,171; green,204; blue,227}] {Instruction Fetch \& Decode};
  \path (nFD.south)+(0.0,-1.3) node (nReorder) [nodeStyle, fill={rgb,255:red,198; green,233; blue,175}] {Reorder Buffer};
  \path (nReorder.south)+(0.0,-1.0) node (nRS) [nodeStyle, fill={rgb,255:red,135; green,220; blue,170}] {Scheduler};
  \path (nRS.south)+(-2.5,0.0) node[anchor=north] (nPort0) [port] {Port 0};
  \path (nRS.south)+(-1.5,0.0) node[anchor=north] (nPort1) [port] {Port 1};
  \path (nRS.south)+(-0.5,0.0) node[anchor=north] (nPort2) [port] {Port 2};
  \path (nRS.south)+(0.5,0.0) node[anchor=north] (nPort3) [port] {Port 3};
  \path (nRS.south)+(1.5,0.0) node[anchor=north] (nPort4) [port] {Port 4};
  \path (nRS.south)+(2.5,0.0) node[anchor=north] (nPort5) [port] {Port 5};
  
  \path (nPort0.south)+(0,-0.75) node[anchor=east] (nPort0FU) [FU] {ALU, V-MUL, \dots};
  \path (nPort1.south)+(0,-0.75) node[anchor=east] (nPort1FU) [FU] {ALU, V-ADD, \dots};
  \path (nPort2.south)+(0,-0.75) node[anchor=east] (nPort2FU) [FU] {Load, AGU};
  \path (nPort3.south)+(0,-0.75) node[anchor=east] (nPort3FU) [FU] {Load, AGU};
  \path (nPort4.south)+(0,-0.75) node[anchor=east] (nPort4FU) [FU] {Store Data};
  \path (nPort5.south)+(0,-0.75) node[anchor=east] (nPort5FU) [FU] {ALU, JMP, \dots};
  
  \begin{pgfonlayer}{background1}    
    \path (nPort0FU.north |- nPort0FU.east)+(-0.25,0.25) node (ee_tl) {};
    \path (nPort5FU.south |- nPort5FU.west)+(+0.25,-0.25) node (ee_br) {};
    \path[fill={rgb,255:red,95; green,211; blue,188}, draw=black!50, rounded corners] (ee_tl) rectangle (ee_br);
  \end{pgfonlayer}
  
  \path let \p{x}=(nRS.south), \p{y}=(ee_br.south) in (\x{x}, \y{y})+(0,-1.0) node (nL1D) 
    [nodeStyle, fill={rgb,255:red,249; green,177; blue,166}] {L1 Data Cache};
  \path (nL1D.south)+(0.0,-1.0) node (nL2) [nodeStyle, fill={rgb,255:red,249; green,177; blue,166}] {L2 Cache};
     
  \draw [->, arrow] (nIC.south) -- (nFD.north);
  \draw [->, arrow] (nFD.south) -- node [right] {4--6 \textnormal\microops} +(0,-0.6) -- (nReorder.north);
  \draw [->, arrow] (nReorder.south) -- node [right] {\textnormal\microops} (nRS.north);
  
  \draw [->, arrow, font=\fontsize{6}{7.2}\sffamily] (nPort0.south) -- node [right] {\textnormal\microop} +(-0,-0.5) -- (nPort0FU.east);
  \draw [->, arrow, font=\fontsize{6}{7.2}\sffamily] (nPort1.south) -- node [right] {\textnormal\microop} +(-0,-0.5) -- (nPort1FU.east);
  \draw [->, arrow, font=\fontsize{6}{7.2}\sffamily] (nPort2.south) -- node [right] {\textnormal\microop} +(-0,-0.5) -- (nPort2FU.east);
  \draw [->, arrow, font=\fontsize{6}{7.2}\sffamily] (nPort3.south) -- node [right] {\textnormal\microop} +(-0,-0.5) -- (nPort3FU.east);
  \draw [->, arrow, font=\fontsize{6}{7.2}\sffamily] (nPort4.south) -- node [right] {\textnormal\microop} +(-0,-0.5) -- (nPort4FU.east);
  \draw [->, arrow, font=\fontsize{6}{7.2}\sffamily] (nPort5.south) -- node [right] {\textnormal\microop} +(-0,-0.5) -- (nPort5FU.east);

  \draw [<->, arrow] (nPort2FU.west) -- (nPort2FU.west |- nL1D.north);
  \draw [<->, arrow] (nPort3FU.west) -- (nPort3FU.west |- nL1D.north);
  \draw [->, arrow] (nPort4FU.west) -- (nPort4FU.west |- nL1D.north);

  \draw [<->, arrow] (nL1D.south) -- (nL2.north);
  \draw [->, arrow] (nL2.east) -- +(+0.8,-0.0) |- (nIC.east);
   
  \background{nIC}{nIC}{nL2}{nFD}{-0.7}{{rgb,255:red,255; green,246; blue,213}}{Front End}
  \background{nReorder}{nReorder}{nL2}{ee_br}{-0.5}{{rgb,255:red,213; green,255; blue,230}}{Execution Engine}
  \background{nL1D}{nL1D}{nL1D}{nL2}{-0.4}{{rgb,255:red,252; green,222; blue,212}}{Memory}
\end{tikzpicture}
\caption{Pipeline of Intel Core CPUs (simplified).}
\labfig{pipeline}
\end{figure}
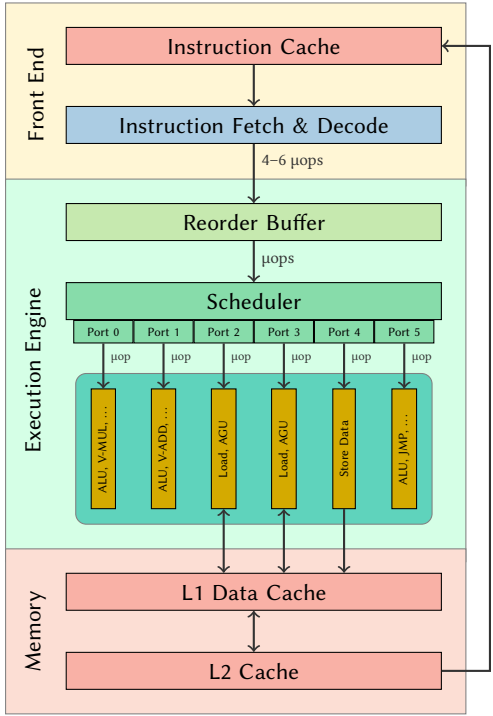

\reffig{pipeline} shows the general structure of the pipeline of Intel Core CPUs.
The pipeline consists of the front end, the execution engine (back end), and the memory subsystem.

The front end is responsible for fetching instructions from the memory, and for decoding them into a sequence of micro-operations~(\microops).

The \emph{reorder buffer} stores the \microops \emph{in order} until they are retired.
It is also responsible for register allocation (i.e., mapping the architectural registers to physical registers), and register renaming (to eliminate false dependencies among \microops).
On some microarchitectures, the reorder buffer can also directly execute certain special \microops, including NOPs, zero idioms (e.g., an \emph{XOR} of a register with itself), and register-to-register moves (\emph{``move elimination''}).

The remaining \microops are then forwarded to the \emph{scheduler} (also known as the \emph{reservation station}), which queues the \microops until all their source operands are ready.
Once the operands of a \microop are ready, it is dispatched through an execution port. Due to \emph{out-of-order} execution, \microops are not necessarily dispatched in program order.
Each port (Intel Core microarchitectures have 6 or 8 of them) is connected to a set of different functional units, such as an ALU, an address-generation unit (AGU), or a unit for vector multiplications.
Each port can accept at most one \microop in every cycle.
However, as most functional units are fully pipelined, a port can typically accept a new \microop in every cycle, even though the corresponding functional unit might not have finished executing a previous \microop.
An exception to this are the divider units, which are not fully pipelined.

\subsection{Assembler Instructions}
Throughout this paper, we will use assembler instructions in Intel syntax.
They have the following form:
\begin{center}
\texttt{mnemonic op$_1$, op$_2$, ...}
\end{center}
The mnemonic identifies the operation, e.g., \texttt{ADD} or \texttt{XOR}.
The first operand \texttt{op$_1$} is typically the destination operand, and the other operands are the source operands (an operand can also be both a source and destination operand).
Operands can be registers, memory locations, or immediates.
Memory operands use the syntax \texttt{[$R_{base}$+$R_{index}$*$scale$+$disp$]}, where $R_{base}$ and $R_{index}$ are general-purpose registers, $disp$ is an integer, and $scale$ is 1, 2, 4, or 8. All of these components are optional and can be omitted.
In addition to these explicit operands, an instruction can also have implicit operands. 

As an example, consider the following instruction:
\begin{center}
\texttt{ADD RAX, [RBX]}
\end{center}
This instruction computes the sum of the general-purpose register \texttt{RAX} and the memory at the address of register \texttt{RBX}, and stores the result in \texttt{RAX}.
We refer to \texttt{RAX} and \texttt{[RBX]} as \emph{explicit} operands.
In addition to that, the instruction updates the status flags (e.g., the carry flag) according to the result. 
The status flags are \emph{implicit} operands of the \texttt{ADD} instruction.

There are often multiple variants of an instruction with different operand types and/or widths.

Note that there is not always a one-to-one correspondence between assembler code and machine code.
Sometimes, there are multiple possible encodings for the same assembler instruction.
It is, in general, not possible to control which of these encodings the assembler selects.
Thus, some machine instructions cannot be generated using assembler code.

\subsection{Hardware Performance Counters}
Hardware performance counters are special registers that store the count of various hardware-related events.
All recent Intel processors have counters for the number of elapsed core cycles, and for the number of \microops that are executed on each port.\looseness=-1

\newcommand{\lat}{\textit{lat}}

\section{Definitions}
In this section, we define the microarchitectural properties we want to infer, i.e., latency,  throughput, and port usage.
\subsection{Latency}
The latency of an instruction is commonly~\cite{intelOptManual17} defined as the ``number of clock cycles that are required for the execution core to complete the execution of all of the \microops that form an instruction'' 
(assuming that there are no other instructions that compete for execution resources).
Thus, it denotes the time from when the operands of the instruction are ready and the instruction can begin execution to when the results of the instruction are ready.

This definition ignores the fact that different operands of an instruction may be read and/or written by different \microops.
Thus, a \microop of an instruction $I$ might already begin execution before all source operands of $I$ are ready, and a subsequent instruction~$I'$ that depends on some (but not all) results of $I$ might begin execution before all results of $I$ have been produced.
To take this into account, we propose the following definition for latency instead.
Let $S = \{s_1, ..., s_m\}$ be the source operands, and $D = \{d_1, ..., d_m\}$ be the destination operands of an instruction.
We define the latency of the instruction to be the mapping $\lat: S\times D \rightarrow \mathbb{N}$ such that $\lat(s_i, d_j)$ denotes the time from when source operand $s_i$ becomes ready until the result $d_j$ is ready (assuming all other dependencies are not on the critical path).
Thus, if $t_{s_i}$ denotes the time at which source operand $s_i$ becomes ready, then destination operand $d_j$ is ready at time $$t_{d_j} = max\{t_{s_i} + lat(s_i, d_j) \mid s_i \in S\}. $$  
With the usual approach of using a single value $\lat$ as the latency of an instruction, this value would be
$$t_{d_j} = max\{t_{s_i}\mid s_i \in S\} + lat,$$
which might be significantly greater than what would be observed in practice.

\subsection{Throughput}
\labelsec{throughput}
When comparing throughput data from different publications, it is important to note that these publications do not all use the same definition of throughput.
Intel defines throughput in its manuals~\cite{intelOptManual12, intelOptManual17} as follows:
\begin{definition}[Throughput - Intel]
\labeldef{throughputIntel}
The number of clock cycles required to wait before the issue ports are free to accept
the same instruction again.
\end{definition}

On the other hand, Agner Fog~\cite{fog17} uses the following definition for (reciprocal) throughput:
\begin{definition}[Throughput - Fog]
\labeldef{throughputFog}
The average number of core clock cycles per instruction for a series of independent instructions of the same kind in the same thread.
\end{definition}
Granlund~\cite{granlund17} uses a similar definition as Fog.

These two definitions are not equivalent, as there can be factors other than contention for the issue ports that may reduce the rate at which instructions can be executed (e.g., the front end, or the memory subsystem).
Moreover, it is not always possible to find instructions of the same kind that are truly independent, as many instructions have implicit dependencies on certain registers or flags.
Hence, the second definition may yield higher throughput values (corresponding to a lower throughput) than the first definition for the same instruction.

Some publications (e.g., \cite{fog17, granlund17}) use the term throughput to denote \textit{instructions per cycle}, while others (e.g., ~\cite{intelOptManual12, intelOptManual17,IACA}) use it to denote \textit{cycles per instruction}.
In this paper, we will use the term with the latter meaning.

\newcommand{\instr}{\textit{instr}\xspace}

\subsection{Port Usage}
Let $P$ be the set of ports of a CPU, and $U$ the set of \microops of an instruction \instr.
Let $ports: U \rightarrow 2^P$ be a mapping such that $ports(u)$ is the set of ports which have a functional unit that can execute the \microop $u$.

We define the port usage $pu: 2^P \rightarrow \mathbb{N}$ of \instr to be a mapping such that $pu(pc) = \abs{\{ u \in U \mid ports(u)=pc \}}$, i.e., $pu(pc)$ denotes the number of \microops of \instr whose functional units are at the ports in $pc$ (we will call the set $pc$ a \emph{port combination}).
Note that, e.g., for a 1-\microop instruction with a \microop $u$ such that $ports(u)=\{0,1\}$, we have that $pu(\{0,1\})=1$, but $pu(\{0\})=pu(\{1\})=0$.

For, e.g., an instruction with $pu(\{0,1,5\})=3$, $pu(\{2,3\})=1$, and $pu(pc)=0$ for all other port combinations $pc$, we will use the notation $3*p015 + 1*p23$ to denote the port usage. 
In other words, the instruction consists of three \microops that may each be executed on ports 0, 1, and 5, and one \microop that may be executed on ports 2 and 3.
\todo{possibly elaborate on this in one sentence}

\section{Algorithms}
\labsec{algorithms}

In this section, we describe the algorithms that we developed to infer the port usage, the latency, and the throughput.

\subsection{Port Usage}
\labelsec{portUsage}
The existing approach by Agner Fog~\cite{fog17} to determine the port usage measures the number of \microops on each port when executing the instruction repeatedly in isolation.
If the result of such a measurement is, e.g., that there is, on average, one \microop on port 0, and one \microop on port 5, the author would conclude that the port usage is $1*p0+1*p5$.

However, this might be incorrect: A port usage of $2*p05$ could lead to exactly the same measurement result when the instruction is run in isolation, but to a very different result when run together with an instruction that can only use port~0 (the \texttt{PBLENDVB} instruction on the Nehalem microarchitecture is an example for such a case).

In another example, if the measurement result is that there are, on average, $0.5$ \microops on each of port 0, 1, 5, and 6, the author would conclude that the port usage is $2*p0156$, whereas the actual usage might be $1*p0156+1*p06$ (this is, e.g., the case for the \texttt{ADC} instruction on the Haswell microarchitecture).

In this section, we propose an algorithm that can automatically infer an accurate model of the port usage.
Our algorithm is based on the notion of a \textit{blocking instruction}: We define a \textit{blocking instruction} for a set of ports $P$ to be an instruction whose \microops can use all the ports in $P$, but no other port that has the same functional unit as a port in $P$.
In the following, we will call the set $P$ a \textit{port combination}.\todo{do we use $P$ later on?}
Blocking instructions are interesting because they can be used to determine whether or not an instruction can only be executed on a given set of ports, the set of ports blocked by the blocking instruction.

Before describing our algorithm to infer a model of the port usage we will now first describe how to find a suitable set of blocking instructions.

\subsubsection{Finding Blocking Instructions}
Let $FU$ be the set of types of functional units that the CPU uses, and let $ports: FU \rightarrow 2^P$ be a mapping from the functional unit types to the set of ports~$P$ that are connected to a functional unit of the given type.
The set of port combinations for which we need to find blocking instructions is the set $\{ports(fu) \mid fu \in FU\}$.

We assume that for each of these port combinations (except for the ports that are connected to the \textit{store data} and \textit{store address} units), there is a 1-\microop instruction that can use exactly the ports in the combination.
This assumption holds on all recent Intel microarchitectures.

Our algorithm first groups all 1-\microop instructions based on the ports they use when run in isolation.
We exclude system instructions, serializing instructions, zero-latency instructions, the \texttt{PAUSE} instruction, and instructions that can change the control flow based on the value of a register.
From the remaining instructions, the algorithm chooses from each group an instruction with the highest throughput (see \refsec{alg:tp}) as the blocking instruction for the port combination corresponding to this group.

As blocking instructions for the port combinations for the ports that are connected to the \textit{store data} and \textit{store address} units, we use the \texttt{MOV} instruction (from a general-purpose register to the memory).
This instruction is a 2-\microop instruction; one of its \microops uses the \textit{store data} unit, and the other a \textit{store address} unit.

To avoid SSE-AVX transition penalties when characterizing SSE or AVX instructions, our algorithm determines two separate sets of blocking instructions for these two types of instructions.
For SSE instructions, the blocking instructions should not contain AVX instructions, and vice versa.


\subsubsection{Port Usage Algorithm}
\todo{achtung, es gab das label portUsage doppelt! habe die zweite Nutzung in der Zeile darŸber kommentiert}

\newcommand{\mymu}{%
  \ifmmode
    \mathchoice
        {\hbox{\normalsize\textmu}}
        {\hbox{\normalsize\textmu}}
        {\hbox{\scriptsize\textmu}}
        {\hbox{\tiny\textmu}}%
  \else
    \textmu
  \fi
}

\LinesNumbered
\newcommand{\assign}{\leftarrow}
\newcommand{\portCombinationsList}{\textit{portCombinationsList}\xspace}
\newcommand{\portCombinations}{\textit{portCombinations}\xspace}
\newcommand{\muopsForCombination}{\textit{{\mymu}opsForCombination}\xspace}
\newcommand{\muops}{\textit{\mymu}ops\xspace}
\newcommand{\blockRep}{\textit{blockRep}\xspace}
\newcommand{\pc}{\textit{pc}\xspace}
\newcommand{\code}{\textit{code}\xspace}
\begin{algorithm}[h!]
\caption{Port Usage}
\labalg{portUsage}
\DontPrintSemicolon
\SetAlgoVlined
$\portCombinationsList \assign$ sort(\portCombinations) \;
$\muopsForCombination \assign []$ //list of pairs\;
\ForEach{\pc\ in \portCombinationsList}{
	$\blockRep \assign 8\cdot$maxLatency(\instr)\;
	$\code \assign $ copy(blockingInstr(\pc), \blockRep)$ ; \instr$\;
	$\muops \assign $ measureUopsOnPorts(\code, \pc)\;
	$\muops \assign \muops - \blockRep$\;
	\ForEach{$(\pc', \muops')$ in \muopsForCombination}{
		\If{$\pc' \subset \pc$}{
			$\muops \assign \muops - \muops'$\;
		}
	}
	\If{$\muops > 0$}{	
		\muopsForCombination.add($(\pc, \muops)$)\;
	}
}
\Return \muopsForCombination
\end{algorithm}

We use \refalg{portUsage} to infer the port usage of an instruction \instr.

The algorithm first sorts the set of port combinations by the size of its elements.
This ensures that, when iterating over the port combinations, combinations that are a subset of another port combination are processed first.

For each port combination \pc, the algorithm determines the number of $\muops$ that may use all of the ports in \pc but no others.
To determine this set, the algorithm concatenates \blockRep copies of the corresponding blocking instruction with the instruction that we want to analyze (line 5).
\mbox{\blockRep} is the product of the maximum latency of the instruction (see \refsec{alg:lat}), i.e., the maximum over the latencies for all input/output pairs, and the maximum number of ports. 
This ensures that there is always a sufficient number of instructions available that can block the ports of the combination.
The operands of the copies of the blocking instructions are chosen such that they are independent from the operands of \instr, and independent from subsequent instances of the blocking instruction.\looseness=-1

Executing the concatenation, instruction \instr will only be executed on one of the blocked ports if there is no other port that it can be executed on.
The algorithm thus measures the number of $\muops$ that use the ports in the combination when executing the code of the concatenation (line 6).
From this value, it subtracts the number of \microops, \blockRep, of the blocking instructions (line 7).
The remaining number of $\muops$ can only be executed on the ports in \pc, otherwise they would have been executed on other ports.

However, it may have been determined previously for a strict subset $\pc'$ of \pc that some or even all of these $\muops$ can only be executed on that subset $\pc'$.
Thus, the number of $\muops'$ on subsets $\pc'$ of the port combination $\pc$, which have been determined in previous iterations of the loop, are subtracted from $\muops$ (line 10).
The remaining number of $\muops$ can be executed on all ports in \pc but on no other ports.

The algorithm can be optimized by first measuring which ports are used when running the instruction in isolation.
The loop then does not need to iterate over all port combinations, but only over those whose ports are also used when the instruction is run in isolation.
Furthermore, we can exit the loop early when the sum of the $\muops$ in the $\muopsForCombination$ list reaches the total number of $\muops$ of the instruction.

\subsection{Latency}
\labsec{alg:lat}
Let $I$ be an instruction with source operands $S$ and destination operands $D$.
We use the following general approach to determine the latency $\lat(s, d)$ for some $s \in S$ and $d \in D$.

Let us first consider the simplest possible case:
\begin{compactenum}
	\item The type of the source operand $s$ is the same as the type of the destination operand $d$.
	\item All instruction operands are \emph{explicit} register operands, and no register operand is both read from \emph{and} written to by~$I$.
\end{compactenum}
Then we can create a \emph{dependency chain} of copies of $I$, such that the register for the destination operand of an instance of $I$ is the register used for the source operand of the next instance of $I$. The other registers should be chosen such that no additional dependencies are introduced.
Given such a chain~$c$ of sufficient length, we can determine $\lat(s, d)$ by measuring the run time of the chain and dividing it by the length of $c$.

Now let us consider the case that the types of the source operand $s$ and the destination operand $d$ are different.
Then it is impossible to create a dependency chain consisting only of instances of~$I$.
To create a chain we need an instruction~$C$ that has a source operand $s_C$ with the same type as $d$, and a destination operand~$d_C$ with the same type as $s$.
We call such an instruction a \emph{chain instruction}.
Given a chain instruction~$C$, we can create a chain by concatenating instances of $I$ and $C$, such that the destination operand of $I$ uses the same register as the source operand of $C$ and vice versa.
Assuming we already know the latency $\lat_C(s_C, d_C)$ of~$C$ we can determine the latency $\lat(s,d)$ by measuring the chain's run time, dividing it by the number of occurrences of $I$, and by subtracting $\lat_C(s_C, d_C)$. 
Chain instructions should ideally be instructions that have as few as possible other operands, and their latency should either be known or easy to determine in isolation.


Now let us assume there are \emph{implicit} operands or register operands that are both read from and written to by~$I$.
Such operands are a challenge as they may introduce \emph{additional} dependencies:
This is the case if $I$ has implicit operands that are both read from and written to (such as, e.g., status flags), or if $s \neq d$ and $s$~or~$d$ are both read from and written to.
If we do not ``break'' these dependencies, then the run time of a chain involving $I$ may be determined by the latency of such an additional dependency, rather than the latency from $s$ to $d$, which we would like to determine.
We break these additional dependencies by adding suitable \emph{dependency-breaking instructions}.
Such instructions overwrite an operand that is part of an unwanted additional dependency, but do not read the same operand.
This makes sure that the run time of the chain is not influenced by dependencies other than the one from $s$ to~$d$.

\paragraph{}
In the following subsections, we describe the most interesting cases of how we create dependency chains for different types of source/destination operands. 

\newcommand{\subsubvspace}{\vspace{2mm}}
\subsubvspace
\subsubsection{Register $\rightarrow$ Register}
\paragraph{Both registers are general-purpose registers}
If both registers are general-purpose registers, we use the \texttt{MOVSX} (``Move with Sign-Extension'') instruction to create a dependency chain.
We do not use the \texttt{MOV} or \texttt{MOVZX} instructions for this purpose, as these can be zero-latency instructions on some microarchitectures in some cases, which can be executed by the reorder buffer (\emph{move elimination}, see~\refsec{pipeline}).
However, \emph{move elimination} is not always successful 
(in our experiments, we found that in a chain consisting of only (dependent) \texttt{MOV} instruction, about one third of the instructions were actually eliminated).
Using the \texttt{MOVSX} instruction avoids this uncertainty.
Moreover, because the \texttt{MOVSX} instruction supports source and destination registers of different sizes, this also avoids problems with \emph{partial register stalls} (see Section 3.5.2.4 of Intel's Optimization Manual~\cite{intelOptManual17}).
A \emph{partial register stall} occurs when an instruction writes an 8 or 16-bit portion of a general-purpose register, and a subsequent instruction reads a larger part of the register.

If at least one of the two register operands is not an implicit operand, we can also use the same register for both operands.
However, if one of the operands is both read and written, it is not possible to add a dependency-breaking instruction for this implicit dependency.
Thus, it would not be possible to analyze the latency of the two operands in isolation in that case.
Moreover, there are some instructions that behave differently if the same register is used for multiple operands.
For example, some instructions with two register operands (like \texttt{XOR} and \texttt{SUB}) are \emph{zero idioms} that always set the register to zero (independent of the actual register content) if the same register is used for both operands.
In all recent microarchitectures, these instructions break the dependency on the register that is used; on some microarchitectures, they can in some cases be executed by the reorder buffer, and do not use any execution ports (see Section 3.5.1.8 of Intel's Optimization Manual~\cite{intelOptManual17}).
There are also other instructions that behave differently on some microarchitectures when the same register is used, for example the \texttt{SHLD} instruction (for details see \refsec{eval:SHLD}).

To be able to detect such a behavior, our algorithm therefore creates microbenchmarks for both scenarios (i.e., using a separate chain instruction, and using the same register for different operands). 

A third option would be to chain the instruction with itself by reversing the order of the two operands (i.e., the destination operand of one instruction would use the same register as the source operand of the next instruction).
However, as this would not work for instructions with implicit register operands, we do not pursue this alternative. 

\paragraph{Both registers are SIMD registers}
Since all MOV instructions for SIMD registers (i.e., XMM, YMM, and ZMM registers) can be zero-latency instructions on some microarchitectures, we use shuffle instructions instead.

SIMD instructions can perform floating-point or integer operations.
If a source for a floating-point operation comes from an integer operation (or vice-versa), a \emph{bypass delay} can occur (see Sections 3.5.1.11 and 3.5.2.3 of the Optimization Manual~\cite{intelOptManual17}).
To capture such cases, we perform measurements with both a floating-point and an integer shuffle instruction as chain instructions.

\paragraph{The registers have different types}
If the registers have different types (e.g., one is a vector register, and the other a general-purpose register), then it is, in general, not possible to find a chain instruction whose latency could be determined in isolation.
Instead, we separately measure and report the execution times for compositions of the instruction with all possible chain instructions with the corresponding types (the number of such instructions is typically rather small).
Note that these times might be higher than the sum of the latencies of the instruction and the chain instruction due to bypass delays.
If we then take the minimum of these times and subtract $1$, we can obtain an upper bound on the latency of the instruction.

\subsubvspace
\subsubsection{Memory $\rightarrow$ Register}
\labelsec{memReg}
To measure the latency of a \texttt{MOV} instruction from the memory to a general-purpose register, we can use a chain of
\begin{center}
\texttt{MOV RAX, [RAX]}
\end{center}
instructions, where we assume that register \texttt{RAX} contains the address of a memory location that stores its own address.
As its address depends on the result of the previous load, the next load can only begin after the previous load has completed.

However, this simple approach would not work for most other instructions, as they usually do not just copy a value from the memory to a register.
Instead, we generalize the approach as follows:
Let $R_a$ be the register that contains the memory address, and $R_d$ be the destination register.
We use
$$\texttt{XOR }R_a\texttt{, }R_d\texttt{; XOR }R_a\texttt{, }R_d$$
to create a dependency from $R_d$ to $R_a$. 
Note that the double \texttt{XOR} effectively leaves $R_a$ unchanged.
However, since \texttt{XOR} also modifies the status flags, we additionally add a dependency-breaking instruction for the flags to the chain.
Furthermore, if $R_d$ is an 8 or 16-bit register, we prepend a \texttt{MOVSX} instruction to the chain to avoid partial register stalls.

The base register of a memory operand is always a general-purpose register.
If the destination register of the instruction is not a general-purpose register, we combine the double \texttt{XOR} technique with the approach for registers described in the previous section to obtain an upper bound on the latency.

\subsubvspace
\subsubsection{Status Flags $\rightarrow$ Register}
As there are no instructions that read a status flag and write a vector register, we only need to consider general-purpose registers here.

To create a dependency from a general-purpose register $R$ to a flag, we use the instruction \texttt{TEST $R$, $R$}.
This instruction reads both register operands (we use for both operands register~$R$), and writes all status flags (except the \texttt{AF} flag).
It has no other dependencies.

\subsubvspace
\subsubsection{Register $\rightarrow$ Memory}
\labelsec{regMem}
It is not directly possible to measure the latency of a store to memory, i.e., the time until the data has been written to the L1 cache.
We can, however, measure the execution time of a chain with a load instruction. 
For the \texttt{MOV} instruction, we could, e.g., measure the execution time of the sequence
\begin{center}
\texttt{MOV [RAX], RBX; MOV RBX, [RAX]}.
\end{center}
However, the execution time of this sequence might be lower than the sum of the times for a load and for a store.
One reason for this is ``store to load forwarding'', i.e., the load obtains the data directly from the store buffer instead of through the cache.
The second reason is that the address of the load does not depend on the preceding store, and thus the address computation might already begin before the store.

While the time of such a sequence does not directly correspond to the latency, it still might provide valuable insights.
We therefore measure the execution time in a chain with a suitable load instruction for all instructions that read a register, and store data to the memory.

\subsubvspace
\subsubsection{Divisions}
For instructions that use the divider units, it is known that their latency depends on the content of their register (and memory, where applicable) operands.
We test these instructions both with values that lead to high latency, and with values that lead to a low latency (we obtained those values from Agner Fog's~\cite{fog17} test scripts).
As most of these instructions use one operand both as input and output operand, and the output of a division with a value that leads to a high latency is often a value that leads to a lower latency, the techniques described in the previous sections for automatically creating dependency chains cannot be used in this case.
We therefore handle these instructions separately.
If, e.g., $R$ is a register that is both a source and a destination register, and $R_c$ contains a value that leads to a high latency, we can use 
$$\texttt{AND }R\texttt{, }R_c\texttt{; OR }R\texttt{, }R_c,$$
or the corresponding vector instructions, to create a dependency chain that always sets $R$ to the same value.

\subsection{Throughput}
\labsec{alg:tp}
As mentioned in \refsec{throughput}, there are different ways of defining throughput.
We will now first describe how we can measure the throughput according to \refdef{throughputFog}.
Then, we will show how the throughput according to \refdef{throughputIntel} can be computed from the port usage.

\subsubvspace
\subsubsection{Measuring Throughput}
To measure the throughput of an instruction, we first try to find a sequence of independent instances of the instruction that avoids read-after-write dependencies as much as possible.
To this end, we select registers and memory locations in a way such that they are not written by one instruction of the sequence and read by a subsequent instruction.
This is, however, not possible for implicit operands that are both read and written.

We then measure the execution time over several repetitions of this sequence, and obtain the throughput by dividing this time by the total number of instruction that have been executed.

We observed that sometimes longer sequences of independent instruction instances can lead to higher execution times per instruction than shorter sequences, in particular, when they use many different memory locations or registers.
We therefore perform measurements for sequences of different lengths (we consider sequences with 1, 2, 4, and 8 elements).

For instructions with implicit operands that are both read and written, we additionally consider sequences with depen\-/dency-breaking instructions.
However, as the depen\-dency-breaking instructions also consume execution resources, this does not necessarily lead to a lower execution time of the sequence in all cases.

For instructions that use the divider units, the throughput can depend on the value of their operands.
We test these instructions both with values that lead to a high throughput, and with values that lead to a low throughput.
For this, we use the same values that we used to measure the latency of such instructions.

\subsubvspace
\subsubsection{Computing Throughput from Port Usage}
Intel's definition of throughput (\refdef{throughputIntel}) assumes that the ports are the only resource that limits the number of instructions that can be executed per cycle, and that there are no implicit dependencies.

If we execute an instruction, for which these requirements hold, repeatedly, then the average wait time until the next instruction can be executed corresponds to the average usage (per instruction) of the port with the highest usage, and the number of \microops on this port will be equal to the execution time (however, this is not true for instructions that use the divider unit, which is not fully pipelined).

For instructions, for which the above requirements do not hold, it is not possible to directly measure the throughput according to Intel's definition.

However, for instructions that do not use the divider unit, it can be computed from the port usage measured in \refsec{portUsage}.
For 1-\microop instructions, the throughput is $\frac{1}{\vert P \vert}$, where~$P$ is the set of ports that the \microop can use.
More generally, the throughput is the solution of the following optimization problem, where $PU$ is the result from~\refalg{portUsage}, and $f(p, pc)$ are variables:
\begin{align*}
&\text{Minimize} & & \max_{p \in Ports} \sum_{(pc,\mu) \in PU} f(p, pc) \\
&\text{Subject to} & & f(p, pc) = 0 & & p \notin pc \\
&&& \sum_{p \in Ports} f(p, pc) = \mu & & (pc, \mu) \in PU
\end{align*}
The variable $f(p, pc)$ captures the share of the \microops that map to the port combination $pc$ that are scheduled on port~$p$.
A schedule maximizing the throughput will minimize the maximum port load $\max_{p \in Ports} \sum_{(pc,\mu) \in PU} f(p, pc)$.

We can reduce this optimization problem to a linear program by replacing the maximum in the objective with a new variable~$z$, and adding constraints of the form $$\sum_{(pc,\mu) \in PU} f(p, pc) \leq z$$ for all $p \in Ports$.
The linear program can be solved using standard LP solvers.

\section{Implementation}
In this section, we describe various aspects of our implementation of the algorithms developed in~\refsec{algorithms}.
\subsection{Details of the x86 Instruction Set}
The algorithms described in~\refsec{algorithms} require detailed information on the x86 instruction set, including, e.g., the types and widths of (implicit and explicit) operands.
While this information is available in Intel's \emph{Software Developer's Manual}~\cite{intelDevManual}, there was, until recently, no sufficiently precise machine-readable description of the instruction set.

Fortunately, Intel recently published the source code of their \emph{x86 Encoder Decoder (XED)}~\cite{XED} library.
The build process of this library uses a set of configuration files that contain a complete description of the x86 instruction set.
While this description is very concise, it is not well documented, and quite complex to parse (collecting the information for a single instruction requires reading multiple files).
It also contains a lot of low-level details, e.g., regarding the encoding, that are not needed for our purposes.

We therefore convert this format to a simpler XML representation that contains enough information for generating assembler code for each instruction variant, and that also includes information on implicit operands.
\subsection{Measurements on the Hardware}
To measure the number of \microops on each port when executing a code sequence, we use hardware performance counters~\cite{intelDevManual}.
For measuring the execution time, we also use a performance counter, as this makes it possible to count core clock cycles, whereas other means to measure the execution time (e.g., using the \texttt{RDTSC} instruction), count reference cycles (which can be different from core clock cycles due to, e.g., frequency scaling).

Before the performance counters can be used, they need to be configured to count the events of interest by writing to a model-specific register.
This requires using privileged instructions.
While it would be possible to read the counters in user mode afterwards, we also perform the measurements in kernel space, as this allows us to also test system instructions.
Moreover, it also allows us to disable preemption and interrupts during the measurements.

To perform one measurement, we generate the following code, where $AsmCode$ consists of $n$ (as explained below) copies of the assembler code sequence we want to analyze.
\negvspace
\begin{algorithm}[h!]
\caption{Measurement}
\DontPrintSemicolon
\SetAlgoVlined
saveState() \label{saveState}\\
disablePreemptionAndInterrupts() \\
serializingInstruction \\
$start \assign$ readPerfCtrs() \label{startPerfCtrs} \\
serializingInstruction \\
$AsmCode$ \label{asmCode}\\
serializingInstruction \\
$end \assign$ readPerfCtrs() \label{stopPerfCtrs}\\
serializingInstruction \\
enablePreemptionAndInterrupts() \\
restoreState() \\
\Return{$end-start$}
\end{algorithm}
\negvspace

The routine saveState() in line~\ref{saveState} saves the content of all registers and flags to the memory, and sets the stack pointer and the base pointer to valid addresses in a large enough memory area that is not used by the main program; this way, the code in line~\ref{asmCode} can freely modify registers or the stack without corrupting the main program. We reserve two registers, however, that store the addresses of the saved state and of the initial performance counter data; the code in line~\ref{asmCode} is not allowed to use these two registers.

We wrap the instructions that read the performance counters (in line~\ref{startPerfCtrs} and~\ref{stopPerfCtrs}) with serializing instructions because otherwise, they could be reordered with the preceding or succeeding code.

The algorithm returns the difference between the two performance counter read operations.
However, this value includes the execution time (or the \microop counts, respectively) of the serializing instructions and (partly) the instructions to read the performance counters, which is undesirable.
We therefore run this algorithm twice. 
The first time, we use $n=10$ copies of the assembler code we want to analyze, and the second time, we use $n=110$ copies. 
By taking the difference of these two measurements, and dividing the result by $100$, we obtain the average run time for one execution of the assembler code sequence we want to analyze.\todo{why not the median?}

We then repeat all these steps $100$ times (after a separate ``warm-up run'' to fill the caches), and finally return the average of the results to minimize the impact of measurement errors.

The values for $n$ were chosen such that the code is small enough to fit in the instruction cache, but large enough to allow for accurate results.


\subsection{Analysis Using Intel IACA}
\labsec{implIACA}
In addition to running the code sequences generated by our algorithms on the actual hardware, we also implemented a variant of our tool that automatically analyzes them with Intel IACA.
IACA treats the code sequences as the body of a loop, and reports average throughput and port usage values for multiple iterations of this loop.
Thus, the results should correspond to the measurements on the actual hardware, which are also averages over executing the code sequences multiple times.\looseness=-1

We consider the IACA versions 2.1, 2.2, 2.3, and 3.0. 
Intel added support for more recent microarchitectures in the newer versions, but at the same time dropped support for older ones.
For microarchitectures that are supported by multiple versions, we analyze the code sequences by all of these versions, as we observed (undocumented) differences between the result from different versions of IACA for the same instructions.

\subsection{Machine-readable Output}
We store the results of our algorithms in a machine-readable XML file.
The file contains the results for all tested microarchitectures, both as measured on the actual hardware, and as obtained from running our microbenchmarks on top of IACA.

\section{Evaluation}\label{sec:evaluation}
In this section, we first describe the platforms on which we ran our tool.

Then, we compare the results we obtained for the port usage by running our microbenchmarks on the actual hardware and by analyzing them with Intel IACA.
We consider a high level of agreement between the two results as evidence that results obtained using performance counter based measurements on microarchitectures which are not supported by IACA are likely also accurate.

Finally, we present several insights that we obtained from the measurement results.
\subsection{Experimental Setup}
\begin{table}
\caption{Tested microarchitectures, number of instruction variants, and comparison with IACA}
\labtab{testedArch}
\negvspace
\negvspace
\begin{center}
  \footnotesize
  \setlength{\tabcolsep}{4.5pt}
  \begin{tabular}{llrcrr}
    \toprule
    Architecture & Processor & \# Instr.& IACA & \microops & Ports \\ 
    \midrule
    Nehalem & Core i5-750 & 1836 & 2.1--2.2& 91.43\% & 95.27\% \\ 
    Westmere & Core i5-650 & 1848 & 2.1--2.2& 91.36\% & 94.61\% \\
    Sandy Bridge & Core i7-2600 & 2538 & 2.1--2.3& 93.25\% & 98.24\%\\
    Ivy Bridge & Core i5-3470 & 2549 & 2.1--2.3 & 91.36\% & 97.39\% \\
    Haswell & Xeon E3-1225 v3\hspace{-.25cm} & 3107 & 2.1--3.0 & 93.10\% & 96.45\%\\
    Broadwell & Core i5-5200U & 3118 & 2.2--3.0 & 92.83\% & 92.64\%\\
    Skylake & Core i7-6500U & 3119 & 2.3--3.0 & 92.29\% & 91.04\% \\
    Kaby Lake & Core i7-7700 & 3119 & - & - & - \\
    Coffee Lake & Core i7-8700K & 3119 & - & - & -  \\
    \bottomrule
  \end{tabular}
\end{center}
\negvspace
\end{table}
We ran our tool on the platforms shown in~\reftab{testedArch}, which includes one processor from each generation of the Intel Core microarchitecture.
The machines have between 4 and 16 GB of RAM.
All experiments were performed using Ubuntu~16.04.
On the Westmere machine, we disabled hyper-threading, as otherwise, the performance counters did not report correct results.

The third column shows the number of instruction variants for which we obtained results. The numbers are higher for newer microarchitectures due to their larger instruction sets.

The total run time of our tool ranges from 50 minutes, on the Coffee Lake system, to 110 minutes, on the Broadwell system.\looseness=-1

\subsection{Hardware Measurements vs. Analysis with IACA}
For the microarchitectures from~\reftab{testedArch} that are also supported by IACA, IACA reports the same \microop count for between 84.65\% (Westmere) and 90.06\% (Broadwell) of the instruction variants that are supported by both tools (we assume that IACA reports the same count if at least one IACA version reports this count; the fourth column in the table shows the IACA versions that support each microarchitecture).
\todo{i think some of this information could also fit into the table!? then you could have absolute numbers + relative numbers; can you give some summarizing data for the nine microarch. in the table?}

If we ignore instruction variants with a \texttt{REP} prefix (which can have a variable number of {\microops}), and instructions with a \texttt{LOCK} prefix (for which IACA in most cases reports a \microop count that is different from our measurements), then the \microop counts are the same for the percentages in the fifth column of~\reftab{testedArch}.

If we consider only the instruction variants for which IACA and our tool report the same \microop count, then in between 91.04\% and 98.24\% of the cases, the port usage as obtained from measurements on the hardware, and as obtained from running our microbenchmarks on top of IACA, is also the same. The percentages for each microarchitecture are shown in the last column of~\reftab{testedArch}. 

\paragraph{Differences Between Hardware Measurements and IACA}
While some of the discrepancies might be due to measurement errors on the hardware, in many cases we were able to conclude that the output of IACA was incorrect.

There are, for instance, several instructions that read from memory, but that do not have a \microop that can use a port with a load unit (e.g., the \texttt{IMUL} instruction on Nehalem).
On the other hand, there are instructions (like the \texttt{TEST mem, R} instruction on Nehalem), that have a \emph{store data} and a \emph{store address} \microop in IACA, even though they do not write to the memory.\looseness=-1

We also found several cases where IACA is not aware that different variants of an instruction have a different port usage.
On the actual hardware, the 32-bit variant of the BSWAP instruction on Skylake, for example, has just one \microop, whereas the 64-bit variant has two \microops. In IACA, both variants have two \microops.

In a number of cases, the sum of the \microops on each of the ports does not add up to the total number of \microops reported by IACA.
An example for this is the \texttt{VHADDPD} instruction on Skylake. 
According to our measurements on the hardware, the port usage of this instruction is $1*p01+2*p5$.
IACA also reports that the instruction has three \microops in total. 
However, the detailed (per port) view only shows one \microop.


\paragraph{Differences Between Different IACA Versions}
We found a number of cases where different IACA versions reported different port usages for the same instructions on the same microarchitecture. 
Often, the results from the newer versions correspond to our measurements on the hardware, so in these cases, the differences seem to be due to fixes of (undocumented) bugs in earlier versions of IACA.
One example for this is the \texttt{VMINPS} instruction on the Skylake microarchitecture. 
In IACA 2.3, this instruction can use the ports 0, 1, and 5, whereas in IACA 3.0 and on the actual hardware, the instruction can only use ports 0 and 1.

On the other hand, we also found a few cases where the results of an older version of IACA correspond to the measurements on the hardware.
An example for this is the \texttt{SAHF} instruction on the Haswell microarchitecture. 
On the actual hardware and in IACA 2.1, this instruction can use the ports 0 and 6. 
In IACA 2.2, 2.3, and 3.0, however, the instruction can additionally use the ports 1 and 5.

\paragraph{Latency/Throughput}
In many cases, it was not possible to obtain accurate latency and throughput data from IACA.
One reason for this is that IACA ignores several dependencies between instructions.
IACA 3.0, for instance, ignores dependencies on status flags; the \texttt{CMC} instruction, for example, which complements the carry flag, is reported to have a throughput of 0.25 cycles by IACA, which is impossible in practice due to the dependency on the carry flag (on the actual hardware, we measured a throughput of 1 cycle).
IACA also completely ignores memory dependencies; the sequence \texttt{mov [RAX], RBX; mov RBX, [RAX]} is reported to have a throughput of 1 cycle.
Furthermore, based on our observations, IACA does not seem to model latency differences between different pairs of input and output operands.\looseness=-1

\subsection{Interesting Results}
\subsubsection{AES Instructions}
We will first look at an example where our new approach for determining the latencies of an instruction revealed undocumented performance differences between successive microarchitectures.

According to Intel's manual~\cite{intelOptManual12}, 
the $\texttt{AESDEC XMM}_1\texttt{, XMM}_2$ instruction has a latency of 8 cycles on the Sandy Bridge architecture.
Agner Fog~\cite{fog17} and AIDA64~\cite{instlatx64} report the same latency. According to IACA 2.1 and the LLVM model, the latency is 7 cycles.\looseness=-1

The instruction reads and writes the first operand, and reads the second operand.
Based on our measurements on the Sandy Bridge system, the latency $\lat(XMM_1, XMM_1)$ is 8 cycles, while $\lat(XMM_2, XMM_1)$ is only about $1.25$ cycles.\todo{how can a latency not be an integer?}
The instruction uses 2 \microops.

According to Intel's instruction set reference~\cite{intelDevManual}, the instruction performs the following operations:
{
\setlength{\interspacetitleruled}{0pt}%
\setlength{\algotitleheightrule}{0pt}%
\begin{algorithm}[h!]
\DontPrintSemicolon
\SetAlgoVlined
\texttt{STATE $\leftarrow XMM_1$ \\
RoundKey $\leftarrow XMM_2$ \\
STATE $\leftarrow$ InvShiftRows(STATE) \\
STATE $\leftarrow$ InvSubBytes(STATE) \\
STATE $\leftarrow$ InvMixColumns(STATE) \\
DEST[127:0] $\leftarrow$ STATE XOR RoundKey
}
\end{algorithm}
}

We can see that the second operand is only needed in the last step (line 6).
So, our latency measurements suggest that one of the two \microops probably computes the XOR operation in the last step (which has a latency of 1 cycle).\todo{can we use this kind of reasoning together with the Intel's instruction set reference to reconstruct the micro ops available on an intel microarchitecture?}

We obtained the same result on the Ivy Bridge system (i.e., Sandy Bridge's successor). On the Haswell system (i.e., Ivy Bridge's successor), on the other hand, the instruction has just one \microop, and the measured latency values for both cases are 7 cycles. The same latency is reported in Intel's manual, by IACA, by the LLVM model, and by Agner Fog.

On the Westmere microarchitecture (i.e., Sandy Bridge's predecessor), which was the first microarchitecture to support the AES instruction set, the instruction has 3 \microops, and we measured a latency of 6 cycles for both operand pairs. 
The same latency is reported in the 2012 version of Intel's manual~\cite{intelOptManual12} (the current version contains no data for Westmere), by IACA~2.1, and by AIDA64.
Agner Fog did not analyze a Westmere system; there is also no LLVM model for Westmere.\todo{what is the purpose of this paragraph?}

We observed the same behavior for the \texttt{AESDECLAST}, \texttt{AES\-ENC}, and \texttt{AESENCLAST} instructions.
To the best of our knowledge, the behavior on Sandy Bridge and Ivy Bridge has not been documented before.

There are also a variants of these instructions where the second operand is a memory operand instead of a register operand.
For these variants, our tool reports for the Sandy Bridge system a latency of 8 cycles for the register-to-register dependency (as before), and an upper bound on the memory-to-register latency of 7 cycles.
According to IACA 2.1 and the LLVM model, the latency is 13 cycles (this value was probably obtained by just adding the load latency to the latency of the register-to-register variants of these instructions).
Agner Fog and AIDA64 do not report the latency of the memory variants.

\subsubvspace
\subsubsection{SHLD} \labsec{eval:SHLD}
We will now see an example that shows that our approach can explain differences among previously published data for the same instruction on the same microarchitecture.

According to the manual~\cite{intelOptManual12}, as well as Granlund~\cite{granlund17}, IACA, and AIDA64, the $\texttt{SHLD R}_1\texttt{, R}_2\texttt{, imm}$ instruction (``double precision shift left'') has a latency of 4~cycles on the Nehalem microarchitecture.
Agner Fog reports a latency of 3 cycles.

The instruction reads and writes the first operand, and reads the second operand.
According to our measurements, $\lat(R_1, R_1)$ is 3 cycles, whereas $\lat(R_2, R_1)$ is 4 cycles.
Thus, $\lat(R_1, R_1)$ corresponds to Fog's result, while $\lat(R_2, R_1)$ corresponds to the latency the other approaches report.

On the Skylake microarchitecture, the same instruction is reported to have a latency of 3 cycles by the manual~\cite{intelOptManual17}, by the LLVM model, and by Agner Fog.
According to Granlund and AIDA64, the latency is 1 cycle.

According to our results for the Skylake system, the latency is 3 cycles if different registers are used for the two operands, but only 1 cycle if the same register is used for both operands (the Nehalem system does not exhibit this behavior).

This suggests that Granlund and AIDA64 test the latency by using the same register for both operands, while Fog uses different registers for both operands, and thus considers only the implicit dependency on the first operand.

\subsubvspace
\subsubsection{MOVQ2DQ}
Next, we will show an example where the port usage is modeled inaccurately by existing work.

According to Agner Fog's instruction tables, the \texttt{MOVQ2DQ} instruction is decoded into 2 \microops on Skylake, one of which uses port 0, while the other can use port 1 and port 5.
This is probably based on the observation that if you execute the instruction repeatedly on its own, then, on average, there is 1~\microop on port 0, and about 0.5 \microops on port 1 and 0.5 \microops on port 5.\looseness=-1

However, our approach shows that the second \microop can actually use port 0, port 1, and port 5.
If we execute the instruction together with a blocking instruction for port 1 and port 5, then all \microops of the \texttt{MOVQ2DQ} instruction will use port~0.\looseness=-1

According to IACA and to the LLVM model, both \microops of this instruction can only use port 5.

\subsubvspace
\subsubsection{MOVDQ2Q}
The following example shows a case where existing work reports an inaccurate port usage on one microarchitecture, and an imprecise usage on another microarchitecture for the same instruction.

On Haswell, the \texttt{MOVDQ2Q} has, according to our results, one \microop on port 5, and one \microop that can use port 0, 1, and~5.
IACA~2.1 reports the same result.
However, according to IACA 2.2, 2.3, 3.0, and the LLVM model, the port usage is $1*p01+1*p015$.
According to Agner Fog, the usage is $1*p01+1*p5$.\todo{so who's right? do we know?}

On Sandy Bridge, our measurements agree with both IACA and the LLVM model for the same instruction ($1*p015+1*p5$).
Agner Fog reports the usage as $2*p015$.

\subsubvspace
\subsubsection{Instructions with Multiple Latencies}
Apart from the examples already described, we also found latency differences for different pairs of input and output operands for a number of other instructions.
This includes most instructions that have a memory operand and another input operand, where such differences can be expected.
We also found differences for the non-memory variants of the following instructions: 
\texttt{ADC},
\texttt{CMOV(N)BE}, 
\texttt{(I)MUL}, 
\texttt{PSHUFB}, 
\texttt{ROL},
\texttt{ROR},
\texttt{SAR},
\texttt{SBB},
\texttt{SHL},
\texttt{SHR},
\texttt{(V)MPSADBW}, 
\mbox{\texttt{VPBLENDV(B/PD/PS)}}, 
\texttt{(V)PSLL(D/Q/W)}, 
\texttt{(V)PSRA(D/W)}, 
\texttt{(V)PSRL(D/Q/W)}, 
\texttt{XADD},\\ and 
\texttt{XCHG}.

For the \texttt{(I)MUL}, \texttt{ROL}, and \texttt{ROR} instructions, this behavior is described in~\cite{intelOptManual17}; for the \texttt{ADC}, and \texttt{SBB} instruction, the behavior has been observed by~\cite{granlund17}.
For the remaining instructions, the differences have, to the best of our knowledge, so far been undocumented. 

\subsubvspace
\subsubsection{Zero Idioms}
According to our results, the \texttt{(V)PCMPGT(B/D/Q/W)} instructions are also dependency-breaking idioms, even though they are not in the list of dependency-breaking idioms in Section 3.5.1.8 of the Optimization Manual~\cite{intelOptManual17}.

\section{Limitations}

Our tool currently has the following limitations:
\begin{compactitem}
\item We only support instructions that can be used in 64-bit mode.
\item We do not support the mostly obsolete x87 floating-point instruction set.
\item A number of system instructions are not supported. This includes, e.g., instructions that write to segment or control registers, instructions that trigger interrupts, the \emph{VT-x} instructions for virtual machines, and instructions that use I/O ports. It also includes instructions like the \emph{undefined instruction} (\texttt{UD}), and the halt instruction (\texttt{HLT}), which cannot be measured in a meaningful way.
\item Except for the division instructions, we do not consider performance differences that might be due to different values in registers, or different immediate values. We do, however, consider immediates of different lengths, e.g., 16 and 32-bit immediates.
\item We do not consider differences due to different memory addressing modes, e.g., with scale and offset. We only test instructions that only use the base register.
\end{compactitem}

\section{Conclusions and Future Work}

We have presented novel algorithms and their implementations to characterize the latency, throughput, and port usage of instructions on all Intel Core microarchitectures, which we believe will prove useful to predict, explain, and optimize the performance of software running on these microarchitectures, e.g., in performance-analysis tools like CQA~\cite{cqa14}, Kerncraft~\cite{hammer15}, or llvm-mca~\cite{llvmmca}.
The experimental evaluation demonstrates that the obtained instruction characterizations are both more accurate and more precise than those obtained by prior work.\todo{feel free to improve with something more expressive}

A machine-readable document with all instruction characterizations generated by our tool is available on our website \url{http://www.uops.info}.
We also plan to release the source code of our tool as open source.
We have also implemented a performance-prediction tool similar to Intel's IACA supporting all Intel Core microarchitectures, exploiting the results obtained in the present work. 

Future work includes adapting our algorithms to AMD x86 CPUs. \todo{add one sentence about what is different there. what needs to be done in addition?}
We would also like  to extend our approach to characterize other undocumented performance-relevant aspects of the pipeline, e.g., regarding micro and macro-fusion, or whether instructions use the simple decoder, the complex decoder, or the Microcode-ROM.

\bibliographystyle{ACM-Reference-Format}
\bibliography{references}

\end{document}